\documentclass{jfm}
\usepackage{graphicx}
\usepackage{epstopdf, epsfig}
\usepackage{amsmath}
\usepackage[dvipsnames]{xcolor}
\usepackage{tikz}
\usepackage{subcaption}
\usepackage[normalem]{ulem}
\usepackage{float}
\usepackage{todonotes}
\usepackage{lipsum}
\usepackage{natbib}
\usepackage{graphicx}
\usepackage{newtxtext}
\usepackage{newtxmath}
\usepackage{natbib}
\usepackage{hyperref}
\makeatletter
\@mparswitchfalse%
\makeatother
\normalmarginpar 

\renewcommand\vec{\mathbf}

\shorttitle{Lubrication-mediated }
\shortauthor{K. A. Phillips, and P. A. Milewski}

\title{Lubrication-mediated rebounds off fluid baths}

\author{K. A. Phillips\aff{1,2}
  \corresp{\email{}},
 \and P. A. Milewski\aff{3}}

\affiliation{\aff{1} Department of Mathematical Sciences, University of Bath,
Bath BA2 7AY, United Kingdom
\aff{2}Mathematics Institute, University of Warwick, Coventry CV4 7AL, United Kingdom
\aff{3} Department of Mathematics, Pennsylvania State University, State College,  PA 16802, USA }

\begin{document}

\maketitle

\begin{abstract}
We present herein the derivation of a lubrication-mediated (LM) quasi-potential model for droplet rebounds off deep liquid baths, assuming the presence of a persistent dynamic air layer which acts as a lubricating pressure transfer. We then present numerical simulations of the LM model for axisymmetric rebounds of solid spheres and compare quantitatively to current results in the literature, including experimental data in the low speed impact regime. In this regime the LM model has the advantage of being far more computationally tractable than DNS and is also able to provide detailed behaviour within the micro-metric thin lubrication region. The LM system  has an interesting mathematical structure, with the lubrication layer providing a free boundary elliptic problem mediating the drop and bath free-boundary evolutionary equations.
\end{abstract}

\begin{keywords}

\end{keywords}

\section{Introduction}
The role of the entrained air layer present throughout droplet rebounds has been well studied. Since the work of \citet{worthington1881impact}, where it was first questioned, it has been well established that rebounds are only able to occur in the presence of a lubricating layer of air preventing the two liquids contacting. The assumption that within the near-contact region, the air can behave like a lubrication layer \citep{hicks2010air} has directed a reach field of literature of impacts on other substrates such as porous media \citep{hicks2017gas}. In the case of the impactor being solid, the same assumption can be made in the pre-impact time; that the layer of air helps delay contact \citep{moore2021introducing}. Recent studies of drops that bounce exist in several configurations, for example off solid hydrophobic surfaces \citep{kolinski2014drops}, and off liquid films \citep{tang2019bouncing}. 

There are fewer studies of drops rebounding off deep baths, particularly in the regime where the underlying fluid is inviscid (or nearly so) and capillary waves are long lived after the droplet impact. These models gained interest in part because of the work of Couder on indefinitely bouncing drops on vibrating baths \citep{couder2005walking} and its fascinating quantum mechanical analogies (\citet{bush2015pilot} and references therein). The studies of  \citet{galeano2017non,galeano2019quasi} and \citet{alventosa2023inertio} imposed a so-called kinematic match between the solid sphere or drop and the bath to avoid modelling the air layer and obtained accurate results when compared to Faraday-pilot wave studies of bouncing and walking droplets. In \citet{galeano2021capillary} the kinematic match model was validated against experiments and DNS simulations of the Navier-Stokes equations of single rebounds of solid hydrophobic spheres for low Weber number impacts.

Current reduced models for solid or droplet rebounds on deep baths do not capture the dynamics of the air layer which is present throughout the drop-bath dynamics. This paper builds off the work on the impact of two-dimensional drops in \citet{Phillips2024modelling}, where the lubrication air layer is considered as an additional fluid region coupling the drop deformation and the wave equations for the bath. We extend that model to three-dimensions in generality (Section 2), and compare its results to the kinematic match, direct numerical simulations and experiments for solid hydrophobic sphere impacts on water (Section 3). 
 

\section{Model}

Consider an almost spherical droplet falling vertically downwards through the air, towards a deep liquid bath initially at rest.  Let the subscripts $\alpha \in \{a,b,d\}$ denote the air, bath, and droplet respectively, and introduce notation $\Omega_\alpha$ for each fluid domain, $\vec{u}_\alpha$ for the fluid velocity, with the density, dynamic viscosity, kinematic viscosity, and pressure of the fluids are given as $\rho_\alpha,\mu_\alpha,\nu_\alpha,p_\alpha$ respectively. Finally we use $\beta \in \{b,d\}$ to denote the bath-air and droplet-air boundaries $\partial \Omega_\beta$, defined by a level set of $F_\beta$ and the surface tension coefficient $\sigma_\beta$ at the air-liquid interfaces.  
In each fluid region we begin from the Navier-Stokes equations, and at the interfaces the usual kinematic condition, stress continuity and velocity continuity boundary conditions $\vec{u}_a  = \vec{u}_\beta$ are taken. Hence, 
\begin{align}
    \nabla \cdot \vec{u}_\alpha &= 0, \qquad &\vec{x} &\in \Omega_\alpha,  \\
    \rho_\alpha ( \partial_t \vec{u}_\alpha + (\vec{u}_\alpha \cdot \nabla) \vec{u}_\alpha) &= - \nabla p_\alpha + \mu_\alpha \nabla^2 \vec{u}_\alpha - \rho_\alpha g \vec{e}_z, \qquad &\vec{x} &\in \Omega_\alpha, \\
    D_t F_\beta &= 0,  \qquad &\vec{x} &\in\partial\Omega_\beta,\\
    p_\beta \vec{n}_\beta + \mu_\beta \tau_\beta \vec{n}_\beta &= p_a\vec{n}_\beta + \mu_a \tau_a \vec{n}_\beta - \sigma_\beta \kappa_\beta, \qquad &\vec{x} &\in\partial\Omega_\beta.
    \end{align}
Where $\tau_\alpha = (\nabla \vec{u}_\alpha + \nabla \vec{u}_\alpha^T)$ is the stress tensor for each fluid, $\kappa_\beta$ denotes curvature of the free surface with outward unit normal vector $\vec{n}_\beta$, gravity $g$ acting downwards in the $-\vec{e}_z$ direction, and $D_t$ is the material derivative. These equations are supplemented by imposing decay of velocities in the far field or appropriate boundary conditions in the case of finite domains. We proceed with several modelling approximations which will reduce the system to a tractable state.

\subsection{Linear quasi-potential model for the liquid bath} 
 
We assume viscous effects are small except in the lubrication layer, and follow methodology from \citet{dias2008theory} which we briefly outline below. Take a Helmholtz decomposition of the liquid as small perturbation from potential flow $\vec{u}_b = \nabla \phi_b + \nabla \times \boldsymbol{\psi_b}$, and use tangential stress conditions and a boundary layer argument to eliminate the $\boldsymbol{\psi}_b$ terms, expressing the rotational part of the flow in terms of $\phi_b$ and $\eta_b$. A more detailed explanation of this argument can be found in \citet{Phillips2024modelling}. 
The free surface of the bath is given by $z=\eta_b(\vec{x},t)$ where $z=0$ is its undisturbed free surface. The governing linear system then becomes
\begin{align}
   0&= \Delta_H \phi_b + \partial_z^2 \phi_b, \qquad &z&\le0,\label{liqBath Laplace}\\
    \partial_t \phi_b &= -\frac{1}{\rho_b}p_a - g\eta_b + \frac{\sigma_b}{\rho_b}\Delta_H \eta_b + 2\nu_b \Delta_H\phi_b, \qquad &z&=0,\\
    \partial_t \eta_b &=  \partial_z \phi_b + 2\nu_b \Delta_H \eta_b, \qquad &z&=0,\label{liqBath KBC}
\end{align}
where $\Delta_H$ is used to denote the horizontal Laplacian, the curvature is small and of the form $\kappa_b = -\nabla_H \eta_b$, and velocity continuity $\vec{u}_b=\vec{u}_a$ is imposed at the interface $\eta_b$ between the bath and air layer. The terms proportional to $\nu_b$ arise from $\boldsymbol{\psi_b}$, and are the corrections to the pressure and vertical velocity, respectively.

\subsection{Air layer lubrication approximation}
Away from the droplet-bath interaction, the air is assumed to be atmospheric pressure and have negligible effect on the system, as we must introduce the domain in which the lubrication effects may be important. The lower part of the droplet is described by  $z=\eta_d(\vec{x}_H,t) = \vec{X}(t)-S(\vec{x}_H,t)$, composed of the shape of the lower part of the droplet $S$, and the vertical height of its centre of mass $\vec{X}$. The domain of $S$ in $\vec{x}_H$ is chosen such that $S$ is single valued and with bounded gradients.  The dynamics of the air layer are described through a lubrication approximation as in \citet{Phillips2024modelling}. Consider a lubrication region $\Omega_a = \Omega_L\times[\eta_b(r,\theta,t),\eta_d(r,\theta,t)]$ where its footprint is described in cylindrical coordinates as 
\begin{equation}
    \Omega_{L} = \{(r,\theta) : r\in [0, r^*(\theta,t)], \theta \in [0,2\pi) \},
\end{equation}
where $r^*$ is the edge of the lubrication region. Denoting $h=\eta_d-\eta_b$, the edge of the lubrication layer is given by a criterion $h(r^*,\theta,t) = \varepsilon$. It is assumed that the $\Omega_L$ is a subset of the domain of S. Assuming that the layer between the droplet and the bath is thin, the leading order balance of the Navier-Stokes equations result in the lubrication balance where $\partial_z p_a = 0$ with 
\begin{align}
    \nabla_H \cdot \vec{u}_a^H + \partial_z w_a &= 0, \qquad &\vec{x}&\in \Omega_{a},\\
        \mu_a \partial_z^2 \vec{u}_a^H &=  \nabla_H p_a, \qquad &\vec{x}&\in \Omega_{a},
\end{align}
and velocity continuity holds at both interfaces: $\vec{u}_a = \vec{u}_d$ at $z =\eta_d$ and $\vec{u}_a = \nabla_H \phi$ at $z =\eta_b$. Here $\nabla_H$ is the horizontal gradient operator, and $\vec{u}_a^H = (u_r,u_\theta)$ is the horizontal velocity of the air layer, such that $\vec{u}_a = (\vec{u}_a^H,w_a)$.
As expected, the pressure $p_a = P(r,\theta,t)$ is independent of $z$ throughout the air layer. Further, we can solve the equation for $\vec{u}_a^H$ in terms of $P$:
\begin{equation}\label{eq: u_H}
    \vec{u}_a^H = \frac{\nabla_H P}{2\mu_a}(z-\eta_b)(z-\eta_d) + \frac{\nabla_H \phi_b|_{\eta_b}}{\eta_b- \eta_d}(z-\eta_d)
    + \frac{\vec{u}^H_d|_{\eta_d}}{\eta_d - \eta_b}(z-\eta_b).
\end{equation}
To obtain the thin film equation we integrate vertically the conservation of mass (incompressibility) equation resulting in
\begin{equation}
    \nabla_H \cdot \int_{\eta_b} ^{\eta_d}  \vec{u}_a^H \;\text{d}z +\left(\vec{u}_a^H \Big|_{\eta_b} \cdot \nabla_H\eta_b -\vec{u}_a^H \Big|_{\eta_d} \cdot \nabla_H\eta_d \right)  + \left( w_a\Big|_{\eta_d} - w_a\Big|_{\eta_b} \right)= 0.
\end{equation}
All terms after the integral can expressed exactly as $\partial_t\eta_d- \partial_t \eta_b = \partial_t h$ from the kinematic conditions at both interfaces. Finally, evaluating the integral in the first term using (\ref{eq: u_H}) to obtain the form of the flux
\begin{equation}
    \vec{Q}= \int_{\eta_b}^{\eta_d} \vec{u}_a^H \;\text{d}z = -\frac{1}{12\mu_a}h^3 \nabla_H P  + \frac{1}{2}h \left(\nabla_H\phi_b + \vec{u}_d^H\right) , \label{Qgeneral}
\end{equation}
results in the thin film equation 
\begin{equation}
    \partial_t h + \nabla_H \cdot \vec{Q} = 0, \qquad \vec{x_H} \in \Omega_{L}.
\end{equation}
As we shall see below this should be considered a nonlinear free boundary elliptic equation for $P$ where $-h_t$ is considered as a forcing term, with the boundary condition $P=0$ at the evolving boundary curve $\partial\Omega_{L}$.

\subsection{Droplet Model}

\subsubsection{Droplet Deformation}

Linear damped droplet oscillations are well understood \citep{lamb1924hydrodynamics} and the results have been used in a variety of contexts e.g. \citet{aalilija2020analytical}. In this section we write equivalent equations for these oscillations using the same quasi-potential theory used above for the bath. Consider a capillary scale droplet of liquid which when unperturbed is a sphere of radius $R_0$. Describing the problem using spherical polar coordinates with azimuthal angle $\theta$ and polar angle $\varphi$, we can take a Helmholtz decomposition for the liquid velocity $\vec{u}_d = \nabla \phi_d + \nabla \times \boldsymbol{\psi}_d$. Using the same argument argument as above \citep{Phillips2024modelling} we obtain similar governing equations to those of the bath \eqref{liqBath Laplace}-\eqref{liqBath KBC} 
\begin{align}
    \Delta \phi_d &= 0, &r<R_0,\\
    \partial_t \phi_d &= -\frac{1}{\rho_d}p_a -  \frac{\sigma_d}{\rho_d} \kappa_d - 2\nu_d \partial_r^2 \phi_d, &r=R_0,\label{DropPhi}\\
    \partial_t \zeta_d &= \partial_r \phi_d  + 2\nu_d\left(\frac{1}{r^2}\Delta_s \zeta_d - \frac{1}{r^3}\int \Delta_s \phi_d \;\text{d}t\right), &r=R_0,\label{DropEta}
\end{align}
where surface of the droplet is given by $r=R_0+\zeta_d(\theta,\varphi)$, and $\kappa_d$ is the mean curvature
\begin{equation} \kappa_d = - \left( \frac{2}{R_0^2} \zeta_d + \frac{1}{R_0^2} \Delta_s \zeta_d \right), \qquad \Delta  = \frac{1}{r^2}\partial_r (r^2 \partial_r) + \frac{1}{r^2}\Delta_s,
\end{equation}
with $\Delta_s$ the Laplace-Beltrami operator (i.e. the surface Laplacian). The equations are valid in the small oscillation limit and have been truncated to contain terms that are linear in the Ohnesorge number (see appendix for a discussion of the nondimensional equations). 
One may proceed with an eigenfunction decomposition of $\zeta_d$ and $\phi_d$ as
\begin{equation}
    \phi_d = \sum_{l=2}^\infty \sum_{m=-l}^l a_{lm}(t) r^l Y_l^m, \qquad \zeta_d =  \sum_{l=2}^\infty \sum_{m=-l}^l c_{lm}(t) Y_l^m, 
\end{equation}
where the $Y_l^m (\theta,\varphi) = e^{im\varphi}P_l^m(cos \theta)$ are spherical harmonics satisfying
\begin{equation}
    -\Delta_s Y_l^m = (l)(l+1)Y_l^m.
\end{equation}
The $P_l^m$ are the associated Legendre polynomials of degree $l$ and order $m$. The terms $l=0,1$ have been omitted as $l=0$ corresponds to a  dilation of the sphere and $l=1$ corresponds to  translational and rotational symmetries which need to be considered separately as they will be affected by the external forces due to the air layer. Substitution into the boundary conditions \eqref{DropPhi}-\eqref{DropEta} leads to the differential equation (where terms of order $Oh^2$ have been omitted)
\begin{equation}
    \ddot{c}_{l,m} + 2\lambda_{l} \dot{c}_{l,m} + \omega_{l}^2c_{l,m} = -\frac{l\hat{p}_{l,m}}{\rho R_0}\;,
\end{equation}
where $\hat{p}_{l,m}$ is the $Y_{l,m}$ coefficient of  of the lubrication air pressure $p_a$. The coefficients of the velocity potential are given to leading order by $a_{l,m}=\frac{\dot{c}_{l,m}}{lR_0^{l-1}}$. The damping coefficients and modal frequencies are given by
\begin{equation}
    \lambda_l = (2l+1)(l-1)\frac{\mu}{\rho R_0^2}, \qquad \omega_l^2 = l(l-1)(l+2) \frac{\sigma}{\rho R_0^3},
\end{equation}
as seen in the literature  \citep{lamb1924hydrodynamics, aalilija2020analytical,  alventosa2023inertio}. A similar equation may be obtained for the coefficients $a_{lm}(t)$, however these are uneccessary for the LM model and have been omitted for brevity. 

\subsubsection{Equations for the centre of mass}
The location of the droplet is determined through updating the position of the centre of mass, given by $\vec{X}(t)$. We shall disregard drag effects from the air, instead only considering contributions from the lubrication layer, then the equation of motion is 
\begin{equation}
    m\ddot{\vec{X}} = F = -\int_{\Omega_L} p_a \; \vec{n_d} \; \text{d}a - mg \; \hat{\vec{z}},
\end{equation}
where $\vec{n_d}$ is the outward normal of the droplet $m = 4\rho \pi R_0^3/3$ is its mass, $g$ is the gravitational constant and $\hat{\vec{z}}$ is the unit vector in the vertical direction. The boundary of the droplet is now given by $\vec{X} + (R_0+\zeta_d(\theta,\varphi)) \vec{\hat{r}}$, where $\vec{\hat{r}}$ is a radial unit vector relative to the position $\vec{X}$.

We note that we have disregarded global rotational motion of the droplet (forced by both asymmetric pressure and shear stresses in the lubrication layer) which would significantly add to the modelling complexity, and which will not be relevant for the vertical (axisymmetric) impacts we shall consider next. We also expect these effects to be small for quasi-normal impacts (e.g. \citet{galeano2019quasi}). 

\subsection{The lubrication-mediated droplet-bath system}

We can now propose a closed system of equations for the evolution of the thin film, liquid bath and droplet. Within the bath and the droplet one may reduce the system further using the framework of Dirichelet to Neumann (DtN) maps. Consider the ``trace" of the potentials on the boundary of the domains in which Laplace's equation are solved
$$\Phi_b(\vec{x_H},t) = \phi_b|_{z=0}, \qquad \Phi_d(\theta,\varphi,t) = \phi_d|_{r=R_0}.$$
The DtN map, denoted $D$, is the linear map that provides the normal derivative of the potential given its value on the boundary. In particular we denote
$$(\partial_z \phi_b)|_{z=0} = D_b \Phi_b, \qquad (\partial_r \phi_d)|_{r=R_0}  = D_d\Phi_d.$$
We shall see that these maps are easily expressed in terms of eigenfunction expansions. Making use of DtN maps we now have the evolution system
\begin{align}
    \partial_t \Phi_b &= -\frac{1}{\rho_b} p_a -g\eta_b+ \frac{\sigma_b}{\rho_b}\Delta_H \eta_b + 2\nu_b \Delta_H\Phi_b , \\
    \partial_t \eta_b &=  D_b \Phi_b + 2\nu_b \Delta_H \eta_b  \doteq F_b, \\
        \partial_t \Phi_d &= -\frac{1}{\rho_d} p_a - \frac{\sigma_d}{\rho_d} \kappa_d - 2\nu_d \partial_r^2 \Phi_d, \\
    \partial_t \zeta_d &= D_d \Phi_d  + 2\nu_d\left(\frac{1}{r^2}\Delta_s \zeta_d - \frac{1}{r^3} \Delta_s D_d^{-1} \zeta_d \right) \doteq F_d ,\\
     m\ddot{\vec{X}} &= F = -\int_{\Omega_L} p_a \; \vec{n_d} \; \text{d}a - mg \; \hat{\vec{z}},\\
    \nabla_H \cdot \vec{Q} &= W+F_d-F_b,  \qquad \vec{x_H} \in \Omega_{L}.\label{EllipticGeneral}
\end{align}

In the surface potential equations, the pressure $p_a$ is nonzero only in $\Omega_L$ and $W=\frac{d\vec{X}}{dt}$ is the vertical velocity of the droplet's centre of mass. In the kinematic condition for the droplet we used the leading order balance $\partial_t \zeta_d = D_d \Phi_d$ to express the right hand side locally in time. While $D_d$ was defined using the velocity potential, its inverse acts on $\zeta_d$ here. This is admissible since it is an invertible linear operator on functions with mean zero, and we may restrict the inverse map to have mean zero also. Equation \eqref{EllipticGeneral}, where $\vec{Q}$ is given by \eqref{Qgeneral} is an inhomogeneous elliptic equation for the pressure with the boundary condition that $p_a=0$ on the free boundary $\partial \Omega_L$. 


\section{Axisymmetric solid impacts and numerical results}

We now consider the simplest case in which this framework can be used and compared to prior simulations and experiments: the rebound of solid (hydrophobic) spheres \citep{galeano2021capillary}. This simplifies considerably the system, reducing the dimension and eliminating droplet oscillations. Considering the axisymmetric domain for the bath to of finite radius $r=L$, and imposing Neumann boundary conditions there for both $\eta_b$ and $\Phi_b$  
we expand these in terms of Bessel functions of the first kind, 
\begin{equation}
    \eta_b(r,t) = \sum_{j=1}^\infty a_j(t) J_0(k_j r), \qquad
    \Phi_b(r,t) = \sum_{j=1}^\infty b_j(t) J_0(k_j r),    
\end{equation}
where the $k_j$ satisfy $J_0^\prime(k_j L) = 0$. We note that in this basis $\phi_b = \sum_{j=1}^\infty b_j J_0(k_j r) e^{k_j z}$ to satisfy Laplace's equation and the DtN map is expressed simply as
\begin{equation}
    D_b \Phi = \partial_z\phi_b|_{z=0} = \sum_{j=1}^\infty k_j b_j J_0(k_j r).
\end{equation}
Hence the DtN operator in this basis corresponds to multiplication by $k_j$. Similarly, the horizontal Laplacian $\Delta_H$ results in a Bessel multiplier $-k_j^2$. The system can now be written for the Bessel coefficients as
\begin{align}
    \label{eq:ODE start}
    \dot{a}_j &= k_jb_j - 2\nu_l k_j^2 a_j, \\
    \dot{b}_j &= -\frac{1}{\rho_b}\hat{p}_j - \frac{\sigma_b}{\rho_b}k_j^2 a_j - 2\nu_l k^2 b_j ,\\
    \label{eq:ODE end} m\ddot{Z} &= {2\pi} \int_0^{r^*} p_a\; r\text{d}r - mg,
\end{align}
with the elliptic equation for pressure
\begin{equation}
    - \frac{1}{r} \partial_r  \left(\frac{r}{12\mu_a} h^3 \partial_r p_a  + \frac{r}{2}h \partial_r \Phi_b \right) = \partial_t h = W- (D_b \Phi_b + 2\nu_l \Delta_H \eta_b), \label{EllipticSimple}
\end{equation}
where $W=\dot{Z}$, $h = Z-\sqrt{R_0^2-r^2} - \eta_b$ for $r\le R_0$, $r^*$ is defined by $h(r^*,t)=\varepsilon$ for $\varepsilon \ll 1$, and where $\hat{p}_j$ are the Bessel coefficients of the Pressure provided by
\begin{equation}
        \hat{p}_j = \frac{2}{(LJ_0(k_jL))^2}\int_0^L p_a(r,t) J_0(k_j r) r \text{d}r\;. \label{eq:BesselCoeff}
\end{equation}
Similar expressions apply for obtaining $a_j$ from $\eta_b$ and $b_j$ from $\Phi_b$. The pressure can be integrated from \eqref{EllipticSimple} using $p_a(r^*,t)=0$ and the symmetry condition $\partial_r p_a(r^*,t)=0$:
\begin{equation}
        \label{eq:pressure} p_a = - \int_{r^*}^r \left[ \frac{ 12 \mu _a}{r'h(r',t)^3} \left(\int _0 ^{r^\prime} r^{\prime \prime} \partial_t h(r'',t) \text{d}r^{\prime \prime} \right) + \frac{ 6 \mu _a}{h(r',t)^2} \partial_r \Phi_b(r',t) \right] \text{d}r^{\prime}
    \;.
\end{equation}

\subsection{Numerical implementation}
We shall compute a solution approximating the system \eqref{eq:ODE start}-\eqref{eq:pressure} by truncating the Bessel expansions at a large $N$ and integrating the resulting differential equations \eqref{eq:ODE start}-\eqref{eq:ODE end} with a fourth order Runge-Kutta scheme. The equation for the pressure \eqref{eq:pressure} is calculated using a trapezium rule. Since Bessel expansions are ill-conditioned on regular grids, we use an oversampled nonuniform grid in $r$ with $M$ points distributed on Chebyshev collocation points on $[0,L]$ using $\theta_j = (j-1)\pi/(M-1)$ with $j = 1\ldots M$ and $r_j = \frac{L(1+\cos(\theta_j))}{2}$.
Hence we compute two matrices: an $M\times N$ matrix which evaluates an N-term Bessel series at $r_j$ and an $N\times M$ matrix corresponding to calculating the projection formula \eqref{eq:BesselCoeff} for the Bessel coefficients of a function given at $r_j$.

\subsection{Model Results}
\begin{table}
    \centering
    \begin{tabular}{l|ccccccc}
        Parameter & Drop & Value & Air  & Value & Bath  & Value&  Units \\
        \hline
        Density   & $\rho_d$ & 250-1200  &   $\rho_a$ & 1.225  & $\rho_b$ & 1000  & kgm$^{-3}$ \\
        Viscosity & $\mu_d$  & 9.78$\times10^{-4}$ & $\mu_a$ & 1.825$\times10^{-5}$ & $\mu_b$  & 9.78$\times10^{-4}$  &  kgm$^{-1}$s$^{-1}$ \\
        Surface tension & - & - &- &- & $\sigma_b$ & 7.2$\times 10^{-2}$ & kgs$^{-2}$ \\
        Radius & $R_0$ & 2.5-8.3$\times10^{-4}$ & -&  -&  -&  -& m  \\
        Initial Velocity & $W_0$ &  0.1-0.45  & -&  -&  -&  -&  ms$^{-1}$ \\ 
    \end{tabular}
    \caption{Parameters used in simulations of a water bath and various solid spheres.}
    \vskip0.1in
    \label{tab: param values}
\end{table}
 We first present results of the rebound of two representative spheres of different radii and densities rebounding off a deep water bath against a sweep of initial downward velocities. The choice of parameters is displayed in table \ref{tab: param values}, and were chosen to correspond to the data for low velocity impacts in \citet{galeano2021capillary}. This will permit the results of the lubrication-mediated model (LM) derived within this paper to be compared to the kinematic match (KM) model \citep{galeano2017non} and direct numerical simulation of the Navier-Stokes equations (DNS). In that study experiments were also performed using hydrophobically coated solid spheres. 

\begin{figure}
    \centering
    \begin{subfigure}{0.32\linewidth}
    \hfill
        \includegraphics[width=\textwidth]{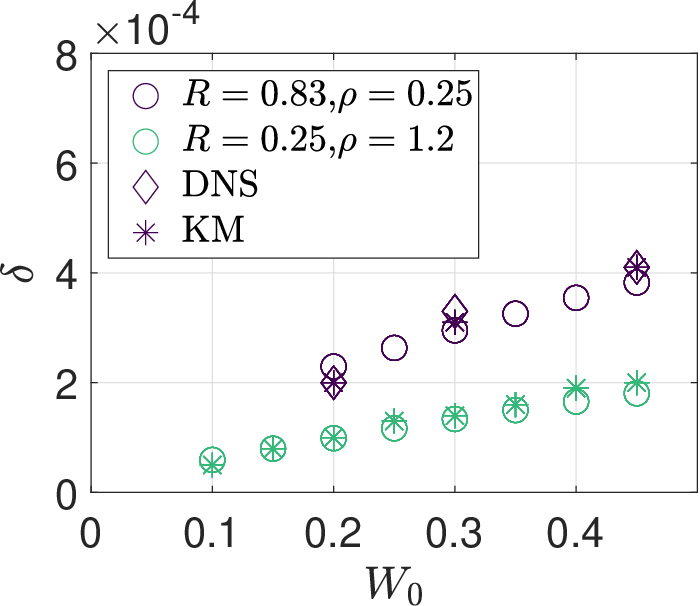}
    \end{subfigure}
    \begin{subfigure}{0.33\linewidth}
        \includegraphics[width=\textwidth]{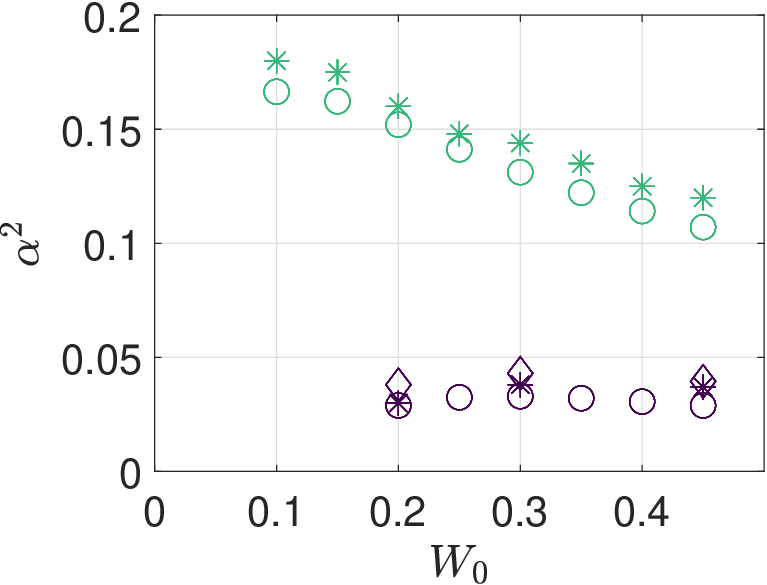}
    \end{subfigure}
    \begin{subfigure}{0.32\linewidth}
    \hfill
        \includegraphics[width=\textwidth]{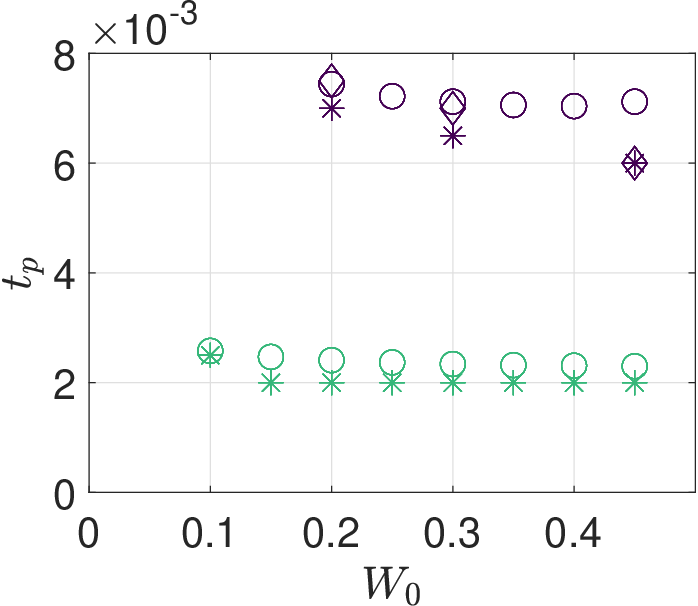}
    \end{subfigure}
    \caption{Comparison of $\alpha^2$, $\delta$, and $t_p$ for two parameter regimes, with LM ($\circ$), compared to DNS ($\Diamond$) and KM ($*$) from \citet{galeano2021capillary}.}
    \label{fig:Param Sweep}
\end{figure}

For such studies, the coefficient of restitution is usually taken as the negative ratio of the speeds between the time of impact $t_\text{imp}$, taken to be when the south pole of the sphere first crosses $z=0$ (correspondingly the centre of mass crosses $z=R_0$), and the time of liftoff $t_\text{lift}$, which is when the sphere leaves the same height on an upwards trajectory. The time spent below $z=0$ is the contact time $t_c = t_\text{lift}-t_\text{imp}$. In \citet{galeano2017non} for low impact velocities, the square coefficient of restitution $\alpha^2$, which is the ratio of mechanical energies before and after impact (using the $z=0$ crossing as reference height) is used, as it can capture very small rebounds where $\alpha^2<0$ where the sphere detaches but it's centre of mass does not reach the $z=R_0$ after rebound. Thus \begin{equation}
    \alpha^2 = \frac{E_{\text{out}}}{E_{\text{in}}} = \frac{W_\text{imp}^2}{W_\text{detach}^2-2g(Z-R_0)},
\end{equation}
where $W_\text{imp}$ and $W_\text{detach}$ are the vertical velocities at the moment of impact $t_\text{imp}$ and separation $t_\text{detach}$, when $\text{min}(h)>\varepsilon$ is first achieved after impact. In Figure \ref{fig:Param Sweep}, $\alpha^2$, the maximum penetration depth $\delta = -(Z_{\text{min}}-R_0)$, and the pressing time $t_p=t_\text{detach}-t_\text{imp}$, are displayed for the LM and KM models and DNS. In \citet{galeano2017non} $t_p$ is favoured over $t_c$ to capture the rebounds which don't return to $z=0$.

Figure \ref{fig:Param Sweep} demonstrates that the agreement is good between the lubrication mediated model against both the results from the KM and DNS, with general trends being upheld. We note here that the latter two methods also involve certain approximations. The KM model assumes no lubrication layer (the method matches the position and velocities of the free surface to those of the sphere) and uses a tangential contact between the sphere and the bath throughout the impact. In addition, for the KM model we are comparing with, the evolution equations for the bath are linear and with the same quasipotential approximation that we use here. The DNS results in \citet{galeano2021capillary} were performed on the open source Gerris  where the solid drop is modelled by using artificially high viscosity and surface tension coefficients. Gerris uses a volume of fluid formulation and interfaces are approximated using continuous density changes (see \citet{popinet2009accurate}).

\begin{table}
    \centering
        \begin{tabular}{c|c c c}
            \textbf{System} & $\alpha$ & $\delta $ & $t_p$ \\
            Lubrication-mediated  & 0.115-0.133 &8.2-8.3$\times 10^{-4}$ m. & 0.0164-0.0168 \\
            Kinematic match& 0.15& 9.5$\times 10^{-4}$m. & 0.0175\\
            Direct numerical simulation &0.24 & 9$\times 10^{-4}$m. & 0.0155\\
            Experimental & 0.12-0.18 & 9-10$\times 10^{-4}$m. & 0.0155-0.0175
        \end{tabular}
    \caption{Results for a falling sphere of radius $R_0= 0.83\times 10^{-4}$ m, density $\rho = 1200$kgm$^{-3}$, and initial velocity $W_0 = -0.35$ms$^{-1}$, onto a deep water bath. The range in the lubrication-mediated data arise from varying bath radius $L$. \\}
    \label{tab: exp comp}
\end{table}

We conclude by providing numerical results for a configuration which can be compared to the KM model, DNS, and also experimental values, as shown in table~\ref{tab: exp comp}. The experimental configuration used a rectangular container while our simulation is cylindrical. In table~\ref{tab: exp comp} we varied the radius of the bath over comparable values to their length and width dimensions resulting in a range of values for our simulations.

The results for the lubrication mediated model restitution coefficient $\alpha$ is at the lower end of the experimental range, the depth of impact is somewhat below the experimental range and the impact time is well within the experimental range. In addition to modelling assumptions, one potential source of error is that the physics of the  hydrophobic coating used in experiments is not considered in any of the models. 
\begin{figure}
    \begin{centering}
        \begin{subfigure}{0.45\textwidth}
            \includegraphics[width=\textwidth]{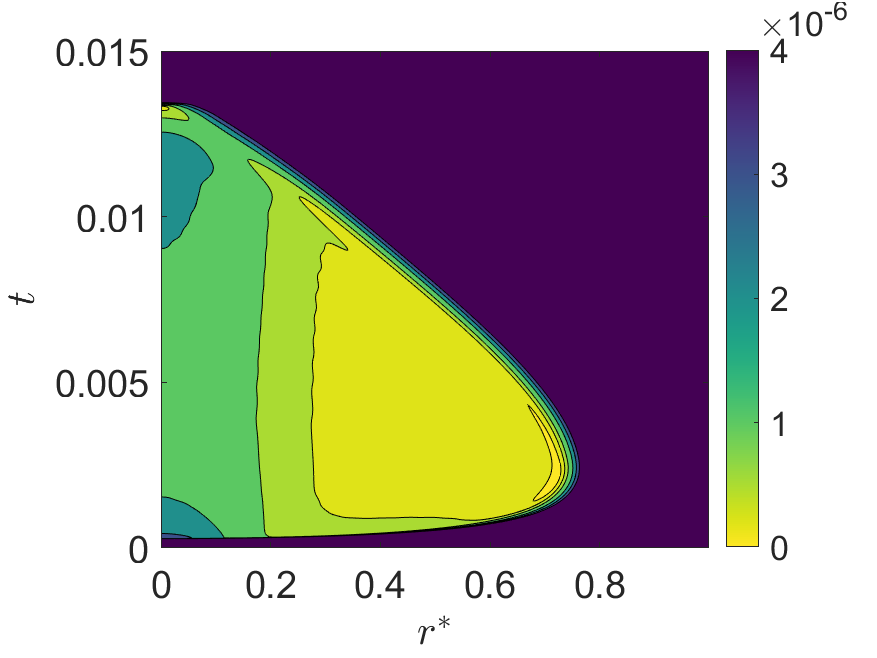} 
           \subcaption{}\label{fig: AirLayer}
        \end{subfigure}
        \begin{subfigure}{0.45\textwidth}
            \includegraphics[width=\textwidth]{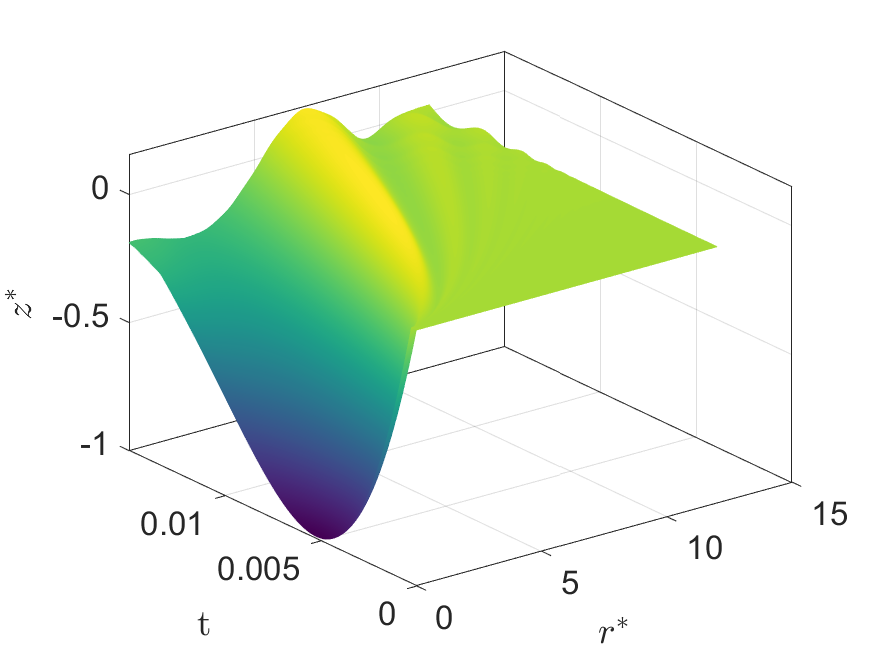} 
            \subcaption{}\label{fig: Eta}
        \end{subfigure}
        \caption{(a) Thickness of the air layer, $h(r^*,t)$. 
        (b) Free surface of the bath, $\eta^*_b(r^*,t)$ shows the sinking depth of the sphere relative to the resultant waves on the free surface.}
\end{centering}
\end{figure}

\begin{figure}
    \begin{centering}
        \begin{subfigure}{0.48\textwidth}
            \includegraphics[width=\textwidth]{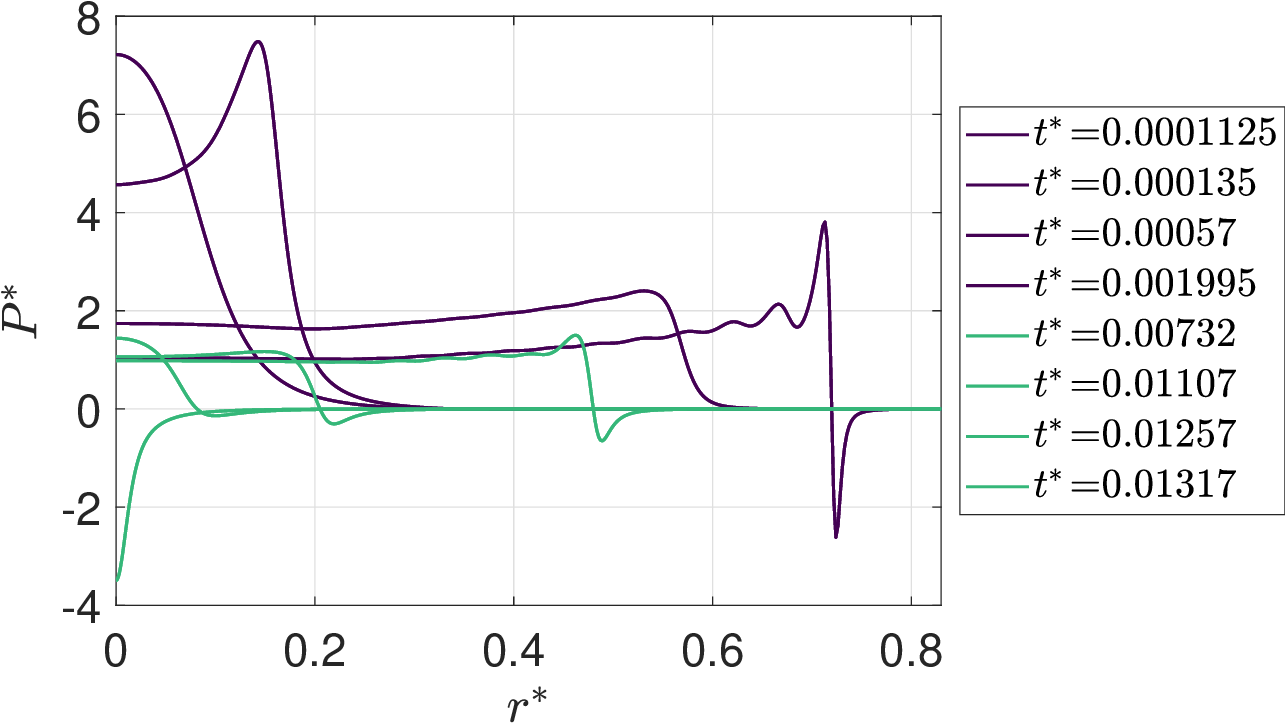} 
            \subcaption{}\label{fig: Pressure Slices}
        \end{subfigure}
        \begin{subfigure}{0.4\textwidth}
            \includegraphics[width=\textwidth]{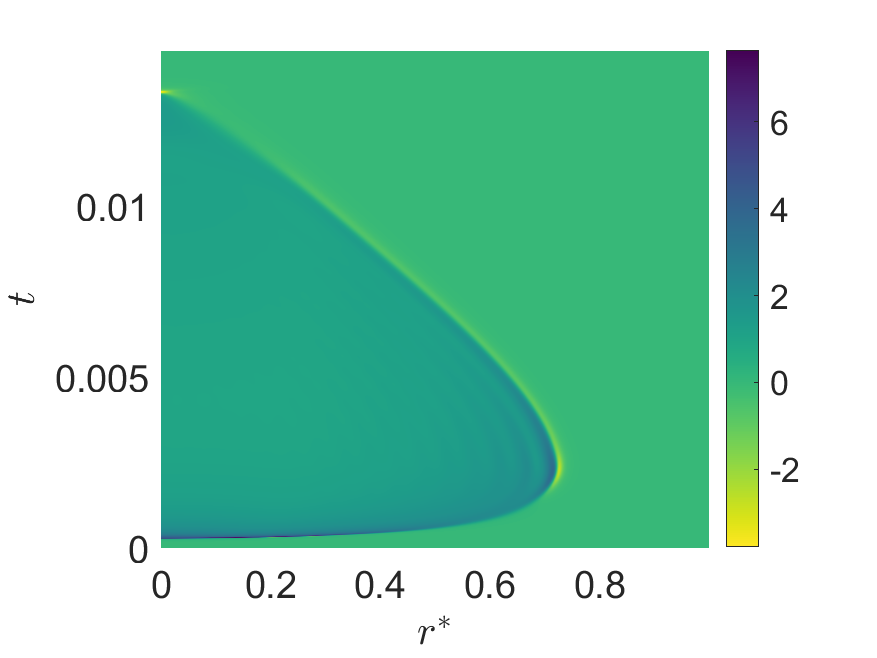} 
            \subcaption{}\label{fig: Pressure}
        \end{subfigure}
        \caption{Lubrication layer pressure (a) Profiles $P^*(r^*,t)$ at selected times. Dark and light curves correspond to expanding and contracting layer respectively. (b) Colormap showing near uniform pressure in the interior and its rapid expansion and slow contraction.}
\end{centering}
\end{figure}

Turning to impact details for this case, Figures \ref{fig: AirLayer} and \ref{fig: Pressure}, show three distinct phases of rebound. First, a rapid expansion of the air layer, driven by the inertia of the impacting drop; second, a slow contraction of the layer; third, a rapid suction event due to the retraction of the free surface. The expansion of the region lasts $\mathcal{O}(1\; \text{ms})$, the contraction lasts $\mathcal{O}(10\; \text{ms})$, and the suction event lasts $\mathcal{O}(1\; \text{ms})$ precedes separation. A highlight of the lubrication mediated model is the ability to capture and quantify the behaviour within the air layer during the rebound. In particular, we can study the evolution of the pressure profile and corresponding height layer within the system, quantities which are challenging to extract even in the DNS setting. In what follows, we have nondimensionalised all lengths with the radius of the sphere (e.g. $r^*=r/R_0$, and area $A^* = A/R_0^2$), the pressure with the surface tension pressure jump across a spherical interface $P^* = P/(2\sigma R_0^{-1})$, and the force with $F^* = F/(2\sigma R_0)$. Throughout the impact, the height of the air layer retains thickness $\mathcal{O}(1)\; \mu$m, as expected from the literature \citep{tang2019bouncing}. However there is some variation as shown in Figure \ref{fig: AirLayer}, with a thickness between 0.5-2 $\mu$m in the bulk of the layer and a narrowing on its boundary to 0.15-0.5$\mu$m. In effect, for the bulk of the impact, the air layer forms a quasi-static cushion constricted on its circular boundary. This narrowing effect is enhanced when the lubrication region boundary ``turns around'' and immediately prior to detachment, where the lubrication layer has shrunk to a small disc at the base of the sphere. Turning to the profile of the pressure within the air layer, shown in figure~\ref{fig: Pressure}, we find a range of 0.9-1.5 $2\sigma R_0^{-1}$ over the vast majority of the region. The maximum pressure is at impact with 7.6 $2\sigma R_0^{-1}$ (indicating an inertially driven pressure) and the minimal pressure being $-$3.8 $2\sigma R_0^{-1}$ at liftoff. A pressure spike occurs where the layer thins at its edge, aligning with results found in \citet{hendrix2016universal}.  Figure~\ref{fig: Pressure Slices} captures the pressure profiles during expansion and contraction. Small oscillations observed in the pressure profile at two intermediate times, we believe are numerical artifacts, as they are lessened by increased mesh resolution. Lesser refined meshes had larger oscillations however with minimal effect on the overall dynamics (restitution coefficient, impact times, generated waves etc...) of the system.

\begin{figure}
    \centering
\includegraphics[width=\textwidth]{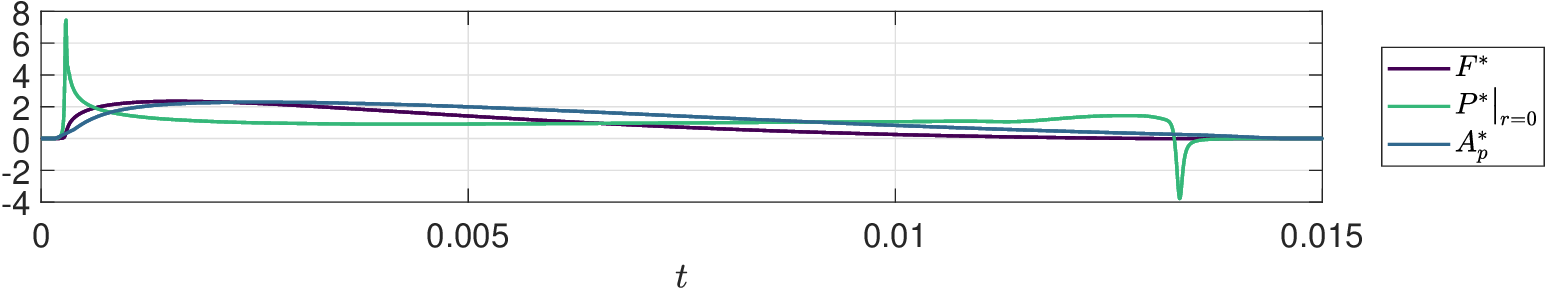}
    \caption{Comparison between the pressure force $F^*$ felt by the droplet, the pressure $P^*$ along $r=0$ during impact, and the pressed area $A_p^*$. Note initial (inertial) and final (suction) spikes. The consequences of the force on the wider system is well tracked by the lubrication layer area.}\label{fig: Force and Pressure}
\end{figure}

Figure \ref{fig: Force and Pressure} shows a temporal slice of pressure, under the south pole of the sphere $P(0,t)$, and the total pressure force acting on the sphere. The pressure has clearly fine scale behaviour at the initial impact and detachment but these features do not substantially affect the overall force on the system. Most of the work on the drop is due to the quasi-uniform pressure at intermediate times. Figure \ref{fig: Eta} shows the full wavefield during impact.

All computations presented were performed in matlab with ranges $N=2^{11}-2^{12}$, $M=2^8-2^9$. For lower impact velocities this was an over-resolution, whereas the finer grids were needed for stronger impacts. The domain length varied $L=12R_0-16R_0$ the latter ensuring the impact is unaffected by waves reflecting from the boundary, even for longer impacts. The longest simulations took $\mathcal{O}(10\; \text{hours})$ on a standard personal computer running Matlab, although the use of an adaptive timestep would probably reduce this considerably. The only free parameter in the system is $\varepsilon$, the cutoff for lubrication and this was set to 40$\mu$m, which corresponds to approximately 5\% the radius of the sphere $R_0$, though values between 10\% and 1\% were also tested and the model displayed insensitivity within this range of $\varepsilon$. The fast decay in pressure in \eqref{eq:pressure} as $h$ increases renders the solution insensitive to larger values of the cutoff $\varepsilon$.

\section{Conclusions}
The lubrication-mediated model herein efficiently resolves certain features of droplet rebounds heretofore unattainable computationally in reduced models. Future work include computational studies including drop deformation and non-axisymmetric impact, coupled with the continued validation through results compared to the wider literature (e.g results from studies such as \citet{tang2018bouncing, alventosa2023inertio} etc). 

This work follows several (e.g. \citet{galeano2019quasi,galeano2017non}) which were motivated by Faraday bouncing droplet dynamics \citep{bush2015pilot}. We expect as future development to use this model with Faraday pilot-wave systems, as an accurate reproduction of the vertical motion of the droplet is thought to be crucial to the observed statistical features resembling quantum mechanics \citep{durey2020faraday}. The present formulation of a lubrication layer mediating the impact of two bodies of fluid also has potential interest in the mathematics community which has recently addressed rigorously questions of whether fluid surfaces can intersect \citep{fefferman2016absence}. \\

\backsection[Declaration of Interests]{The authors report no conflict of interest.}

\backsection[Funding]{This study has been supported by Engineering and Physical
Sciences Research Council project no. EP/S022945/1 (K.A.P), and  P.A.M gratefully acknowledges support through the Leverhulme project RPG-2023-264.}
\appendix
\section{}

For simplicity, assuming the drop and bath are the same liquid, the equations can be nondimensionalised using length scale $R$ and surface-tension-based pressure, time, and velocity scales $\sigma/R_0$, $(\rho_b R_0^3/\sigma)^{1/2}$ and $(\sigma/\rho_b R_0)^{1/2}$ respectively. 
The resulting system is
\begin{align}
    \partial_t \Phi_b &= - p_a -\text{Bo} \,\eta_b + \Delta_H \eta_b + 2 \,\text{Oh} \, \Delta_H\Phi_b , \\
    \partial_t \eta_b &=  D_b \Phi_b + 2 \,\text{Oh}\, \Delta_H \eta_b  \doteq F_b, \\
    \partial_t \Phi_d &= - p_a - \kappa_d - 2 \,\text{Oh} \, \partial_r^2 \Phi_d, \\
    \partial_t \zeta_d &= D_d \Phi_d  + 2 \, \text{Oh} \left(\frac{1}{r^2}\Delta_s \zeta_d - \frac{1}{r^3} \Delta_s D_d^{-1} \zeta_d \right) \doteq F_d ,\\
    \ddot{\vec{X}} &= F = -\int_{\Omega_L} p_a \; \vec{n_d} \; \text{d}a - Bo \; \hat{\vec{z}},\\
    \nabla_H \cdot \vec{Q} &= W+F_d-F_b,  \qquad \vec{x_H} \in \Omega_{L}.
\end{align}
where all variables are dimensionless and the Bond number $Bo=\rho_b g R_0^2/\sigma$ and the Ohnesorge number $Oh = \mu_b/\sqrt{\rho_b R_0\sigma}$ appear. The derivation of these equations assume small surface deformations (resulting in linear wave systems) and smallness of $Oh$. Indeed the system has already been truncated at leading order in $Oh$ when applying stress balances \citep{milewski2015faraday}. Consequently, we may therefore formally eliminate any terms $Oh^2$ arising from further manipulation of the equations. For the lubrication layer, using a layer thickness $\varepsilon$, we obtain
\begin{equation}
    \vec{Q}=  -\frac{\varepsilon^3}{R_0^3} \left(\frac{\mu_a}{\sqrt{\rho_b R_0 \sigma}}\right)^{-1} \frac{1}{12} h^3 \nabla_H P  + \frac{\varepsilon}{R_0} \frac{1}{2}h \left(\nabla_H\phi_b + \vec{u}_d^H\right) , 
\end{equation}
hence the balance $\varepsilon/R_0 \sim (\mu_a/\sqrt{\rho_b R_0 \sigma})^{1/3}$, which is a similar law to that found in \citet{moore2021introducing}, $\varepsilon/R_0 \sim (\mu_a/\rho_b R_0 W_0)^{1/3}$ for inertially-scaled cushioning, where $W_0$ is the impact speed. In our case $\varepsilon/R_0 \approx 0.05$ the value we have used in simulations.

\bibliographystyle{jfm}

\bibliography{bibliography}

\end{document}